\documentclass[12pt]{iopart}

\usepackage{fixltx2e}
\usepackage{fullpage}
\usepackage[utf8]{inputenc}
\usepackage[T1]{fontenc}
\usepackage[english]{babel} 
\usepackage{graphicx}
\usepackage{textcomp}
\usepackage{mathpazo}
\usepackage{subfigure}
\usepackage{rotating}
\usepackage[compact]{titlesec}
\usepackage{cite}

\DeclareGraphicsExtensions{.pdf,.png,.jpg,.eps} 

\newcommand{\pip}{$\pi^{+}$}
\newcommand{\pim}{$\pi^{-}$}

\newcommand{\pT}{\ensuremath{p_{\mathrm {T}}}}
\newcommand{\dedx}{\ensuremath{\mathrm {d}E/\mathrm {d}x}}
\newcommand{\sdedx}{\ensuremath{\sigma_{\mathrm{d}E/\mathrm{d}x}}}
\newcommand{\mdedx}{$\langle$d$E$/d$x\rangle$}

\newcommand{\Dpi}{\ensuremath{\Delta_{\pi}}} 
 
\newcommand{\mdedxpi}{$\langle$d$E$/d$x\rangle_{\pi}$}

\newcommand{\gev}{GeV$/c$}
\newcommand{\bg}{$\beta\gamma$}

\begin{document}
\title{Charged pion identification at high \pT~in ALICE using TPC \dedx}
\author{L. Bryngemark (for the ALICE Collaboration)}
\address{Department of Physics, Division of Experimental High Energy Physics,
  Lund University, Box 118, S-221 00 Lund, Sweden}
\ead{Lene.Bryngemark@hep.lu.se}

\begin{abstract}
The ALICE TPC provides excellent charged particle tracking for the study of pp
and Pb–Pb collisions at LHC. The TPC also allows particle identification via the
measurement of the specific ionisation \dedx. At high \pT~(\pT~> 3 \gev) this
is accomplished in the region of the relativistic rise of the energy
loss. From the energy loss distributions the yields of charged pions, kaons, and protons can be determined in bins of \pT~and \pT-spectra are constructed.
Here we present the performance of such an analysis in pp collisions at 7 TeV for charged pions up to 10 \gev.
\end{abstract}

\section{Introduction}
Flow measurements and particle spectra from ultrarelativistic heavy ion
collisions at RHIC and LHC have so far given many interesting results in the
intermediate and high transverse momentum region $\pT>2$~\gev. In particular,
measurements of elliptic flow \cite{flow} and high-\pT~suppression for
identified particles \cite{ptSupp} at RHIC have given insight in hadronisation
mechanisms and the evolution of the quark gluon plasma. 

This report presents a method for identifying charged particles at high
\pT~with the TPC in ALICE. Performance results are shown for $3\leq\pT\leq 10$~\gev.
\section{The ALICE TPC}
The ALICE Time Projection Chamber (TPC) is a gaseous tracking detector with
full azimuthal acceptance, covering a pseudorapidity range of $|\eta|<0.9$ for
full track length within the TPC volume \cite{TPCpaper}. It is a large (90
m$^3$ gas volume) cylindrical barrel, with read-out at the two end caps. The
drift field of 400 V/cm is generated by a central HV cathode.
The end caps are segmented into 18 trapezoidal sectors each with Inner and
Outer Read-Out Chambers (IROCs and OROCs), equipped with MultiWire Proportional
Chambers (MWPCs). Here the signals are read out on 159 pad rows in the radial
direction, for a total of about 560 000 pads. Such high read-out granularity
is essential for tracking in the high-multiplicity environment of central
heavy ion collisions, which is what ALICE is designed for. The front end
electronics have on-board digital filters, allowing baseline restoration, 
cancellation of signal tails due to ion drift and data reduction via
``zero-suppression''.

\begin{figure}[!h]
  \begin{center} 				
    \includegraphics[width=0.55\linewidth]{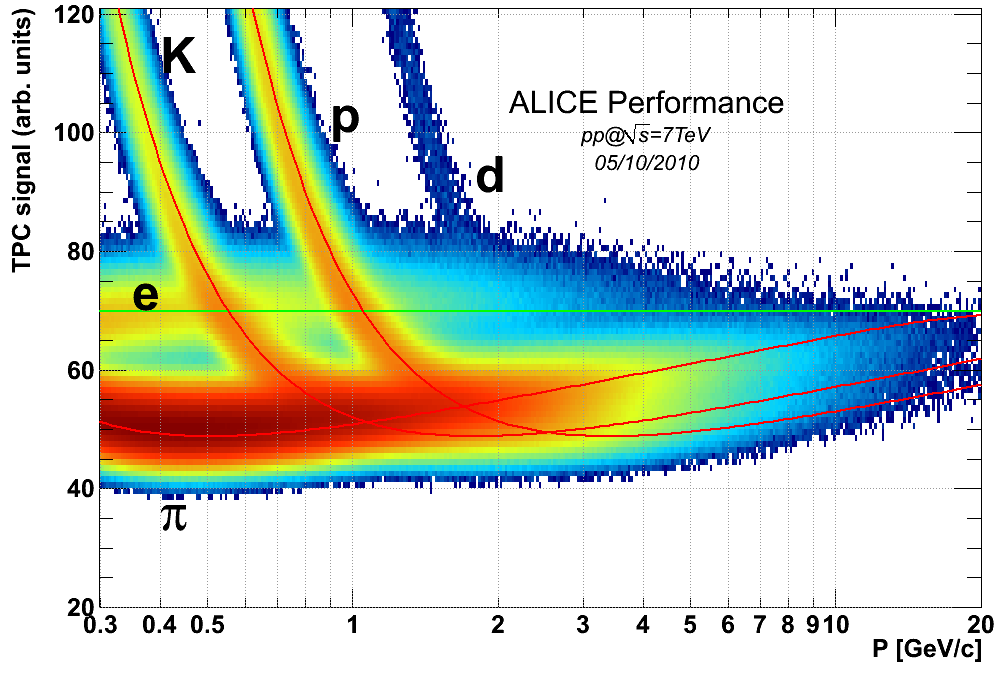}
    \caption{(Colour online) TPC signal as function of momentum in pp collisions at 7 TeV. Pions can be clearly separated at their minimum energy loss around $p=0.5$~\gev.}
    \label{fig:TPCdedx}
  \end{center}
\end{figure}

The TPC is used for tracking and measuring charged particle energy loss
(\dedx) and momentum $p$, simultaneously. The latter two can be combined and
used for particle identification (PID), as energy loss for a given charge
follows a single curve in \bg~$=\frac{p}{m}$, $m$ being the particle
mass. This curve can be well described by a Bethe-Bloch parametrisation, and the
energy loss of each particle species will follow its own curve in $p$. Figure
\ref{fig:TPCdedx} illustrates energy loss vs $p$ as measured by the ALICE
TPC. Bethe-Bloch parametrisation curves for the different particle 
species are also drawn in the figure. The charge deposited on read-out pads
along a track (charge clusters) follows a Landau-like distribution, with a
tail of few instances of high-energy transfer. For this reason the energy loss
is calculated as a truncated mean (the lowest 60\%) of the distribution of
track cluster charge. This is referred to as TPC signal in
Fig. \ref{fig:TPCdedx}, or simply \dedx~in the following. 

The ALICE TPC \dedx~resolution ($\frac{\sdedx}{\dedx}$) is better than
5\% for full length tracks, and the \pT~resolution $\Delta\pT/\pT$ as of
December 2009 was $\sim7\%$ at 10~\gev.

As is seen in this figure, in the low momentum region of a \dedx~vs $p$
histogram, the particle species line up in bands according to their mass and
charge. As \dedx~and $p$ are both determined from the TPC there is no
possibility for mismatches.

On the relativistic rise, $p>3$~\gev, the energy loss curves follow a
logarithmically rising behaviour. Here the particle species
curves are not separated enough for a unique particle identification on a
track-by-track basis. Instead a statistical particle identification for the
integrated yields is employed. This method will be described in the following.

\section{Method}

Event and track selection is carried out following the ALICE unidentified
charged hadron \pT~spectra analysis \cite{pt900}. An additional cut is
applied to reject tracks crossing areas between the TPC read-out
sectors which are not instrumented, since these tracks have worse
\dedx~resolution, due to the smaller number of ionisation measurements. This
analysis focuses on high-momentum tracks, which are only bent a little by 
the magnetic field. With increasing rigidity, only tracks with an original
angle pointing to the area between two sectors will be significantly affected.

The method is based on the energy loss, which is parametrised in $p$.
Final results are obtained from fits in \pT. The main steps of the method are
as follows: 
\begin{enumerate}
\item A 2D histogram is filled with \dedx~vs~$p$, measured for each track in the TPC.
\item \label{item:constraints} This 2D histogram is fitted with a sum of three
  Gaussians ($\pi$, K, p) for each $p$ bin, where the mean of each Gaussian
  follows a common parametrisation of the Bethe-Bloch curve. The \mdedx~vs
  \bg~dependence is thus extracted in one simultaneous fit. The plateau is
  extracted using a clean electron sample in the low-\pT~region. 
\item For each \pT~interval, a 1D histogram is filled with $\Dpi \equiv \dedx - $\mdedxpi~(for each track).
\item The \Dpi~histograms are fitted with a sum of four Gaussians: one for
  each of the particle species p, K, $\pi$ and e. The yields are the only free
  parameters; the others (8 out of 12) are fixed in the following way:
  \begin{itemize}
  \item [-] The Bethe-Bloch fit extracted in (\ref{item:constraints}) is used to
    fix the means.
  \item [-] The widths of the Gaussians are fixed to values determined using a
    clean sample of minimum ionising pions, from the observed relation that the
    relative width is found to be constant (the width scales with \mdedx).
  \end{itemize}
\item From the fits, particle yields as a function of transverse momentum are
  determined.
\end{enumerate}
\begin{figure}[h]
  \begin{center}
    \subfigure{\label{fig:fitPt1}\includegraphics[width=0.48\linewidth]{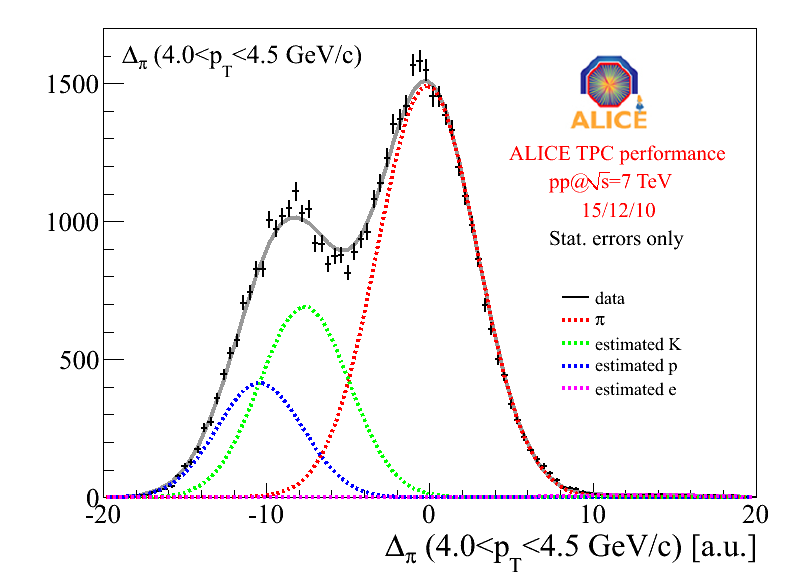}}
    \hspace{0.3cm}
    \subfigure{\label{fig:fitPt2}\includegraphics[width=0.48\linewidth]{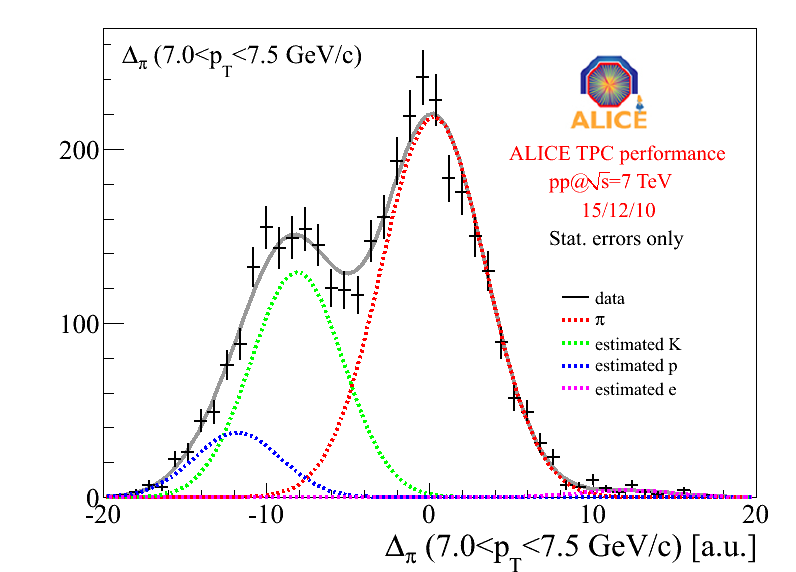}}
    \\
  \end{center}
  \caption{(Colour online) Examples of the \Dpi~distribution in two \pT~bins,
    fitted with a sum (solid grey line) of four Gaussians (for p, K, $\pi$ and
    e, dotted coloured lines). The electron contribution is negligibly small.}
  \label{fig:fits}
\end{figure}
Examples of fits are shown for two \pT~bins in Fig. \ref{fig:fits}. 
Positive and negative particles are treated both together, to increase
statistics, and separately, to enable the study of antiparticle-to-particle
ratios (expected to be close to 1 at mid-rapidity at LHC).

\section{Performance of the \dedx~analysis}
As was seen in the two example fits above, the fit to the $\pi$ peak is
completely determined from the right-hand side, making these fit results very
stable (since it is essentially a single Gaussian fit). 
Protons and kaons have smaller mass ratio than kaons and pions, giving smaller
separation. The quality of the fit result for these species is thus harder to
verify. For this reason, the remainder of this report focuses on the pions,
even though identification of all three species is pursued. 

\begin{figure}[h]
\begin{center}
\subfigure[\pip~to \pim~raw yield ratio, as a function of \pT.]{\label{fig:pmRatio}\includegraphics[width=0.48\linewidth]{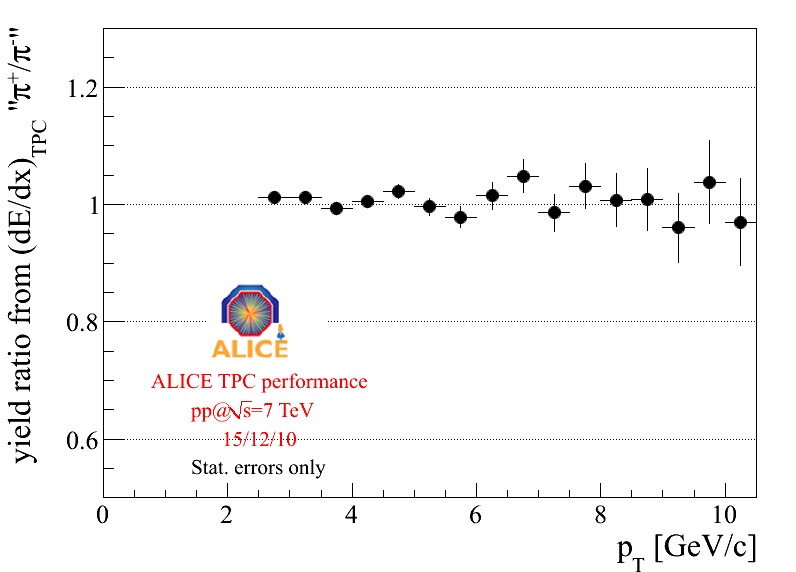}}
\hspace{0.3cm}
\subfigure[(Colour online) Estimated \pip~and \pim~fraction out of the yield of
      all charged hadrons, as a function of \pT.]{\label{fig:piRatio}\includegraphics[width=0.48\linewidth]{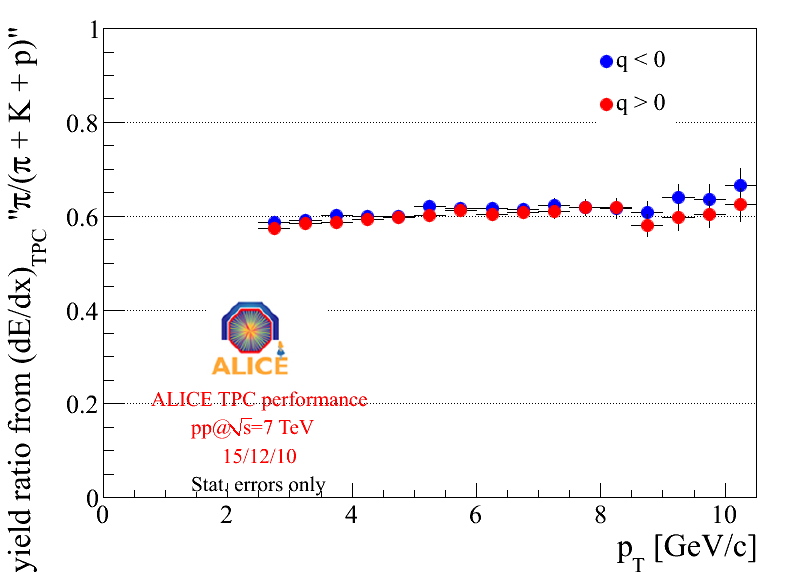}}
\\
  \end{center}
  \caption{Uncorrected results as obtained from fits to the
    \Dpi~distributions.}
  \label{fig:ratios}
\end{figure}

\begin{figure}[h]
  \begin{center}
    \includegraphics[width=0.45\linewidth]{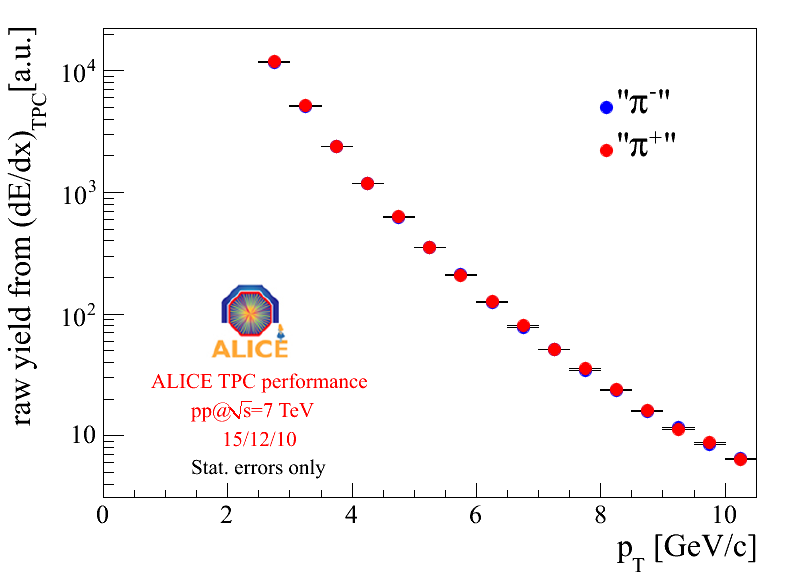}
    \caption{(Colour online) Uncorrected charged pion \pT~spectrum.}
    \label{fig:ptSpectrum}
  \end{center}
\end{figure}

Apart from visually inspecting the fit results, a number of crosschecks are
done to assess the performance of the method. This is illustrated in
Fig. \ref{fig:ratios}. One consistency check is to take
the ratio of the estimated positive and negative pion yields. The raw yield
ratio shown in Fig. \ref{fig:pmRatio} is consistent with the expected
value of 1. Figure \ref{fig:piRatio} shows the resulting raw $\pi$ yield
fraction out of the total hadron yield, as a function of \pT. Here it is seen
that the obtained yield fractions of positive and negative pions,
respectively, agree within statistical errors over the full \pT~range. 
The raw charged pion yield as a function of \pT~obtained with this method is
shown in Fig. \ref{fig:ptSpectrum}, without normalisation or other corrections. 

\section{Conclusion}
A method using the TPC \dedx~for identification and yield extraction of charged
particles at high \pT~has been presented. The very good
performance of the ALICE TPC, and the stable performance results of the method
presented here, are very promising for extracting in particular the charged
pion yields.\\

\section*{References}

\bibliographystyle{unsrt}
\bibliography{BryngemarkProceeding_arXiv}
\end{document}